\documentclass{jaa}
\usepackage{natbib}
\usepackage{graphicx}

\newcommand{\apj}{ApJ}
\newcommand{\apjs}{ApJS}
\newcommand{\aj}{AJ}
\newcommand{\aap}{A\&A}
\newcommand{\mnras}{MNRAS}
\newcommand{\nat}{Nature}
\newcommand{\pasp}{PASP}
\newcommand{\araa}{ARA\&A}
\newcommand{\apjl}{ApJL}
\newcommand{\prd}{PRD}
\newcommand{\aplett}{ApL}
\newcommand{\pasj}{PASJ}
\newcommand{\pasa}{PASA}

\usepackage{hyperref}
\hypersetup{
  colorlinks   = true, 
  urlcolor     = blue, 
  linkcolor    = blue, 
  citecolor   = blue 
}

\begin{document}\sloppy

\title{A chemodynamical study of $r$-process-enhanced stars}




\author{Pallavi Saraf \textsuperscript{1,2}, 
Thirupathi Sivarani\textsuperscript{2}, Carlos Allende Prieto\textsuperscript{3,4}, 
Shashikiran Ganesh\textsuperscript{1},
 and Drisya Karinkuzhi\textsuperscript{5}}
\affilOne{\textsuperscript{1}Physical Research Laboratory, Ahmedabad, India}

\affilTwo{\textsuperscript{2}Indian Institute of Astrophysics, Koramangala 2nd Block, Bangalore, 560034, India\\}

\affilThree{\textsuperscript{3}Instituto de Astrof\'isica de Canarias, E-38200 La Laguna, Tenerife, Spain\\}

\affilFour{\textsuperscript{4}Departamento de Astrof\'isica, Universidad de La Laguna, E-38205 La Laguna, Tenerife, Spain\\}

\affilFive{\textsuperscript{5}Calicut Universiy, Kerala\\}


\twocolumn[{

\maketitle

\corres{pallavi@prl.res.in}

\msinfo{}{}

\begin{abstract}
The $r$-process enrichment in the Galaxy still remains elusive with regard to its nucleosynthesis conditions and the astrophysical sites where it occurs. As part of ongoing efforts to pinpoint the origin of chemically peculiar $r$-process-enhanced (RPE) stars, we concentrate in this study on the kinematics of RPE stars to investigate possible variations in the $r$-process enrichment among the Galactic components. We calculate the orbital parameters of a sample of 472 metal-poor RPE stars and associate them to the Galactic bulge, disk and halo populations using a physically motivated classification based on apocenter distance and maximum absolute vertical height of the orbit. We show that the Toomre diagram does not properly separate stars in the disk and halo components when they are on highly eccentric and/or retrograde orbits. The Galactic disk and halo share a similar fraction of RPE stars, in contrast to the earlier perception that the majority of RPE stars belong to the halo. We find that the stars most likely to be accreted belong to the halo. However, 3/4 of the stars lie in a mixed-zone. The inner disk, inner halo and outer halo stars exhibit similar abundance trends for the n-capture elements.
\end{abstract}


}]


\doinum{12.3456/s78910-011-012-3}
\artcitid{\#\#\#\#}
\volnum{000}
\year{2023}
\pgrange{1--9}
\setcounter{page}{1}
\lp{9}

\section{Introduction}
\label{sec:introduction}
The galactic chemical evolution (GCE) of the Milky Way and other galaxies is far from being well understood, especially when it comes to the enrichment of heavy elements. The nucleosynthesis of elements heavier than Zn is primarily governed by two neutron capture processes, namely, the slow neutron capture process ($s$-process) and the rapid neutron capture process ($r$-process) \citep{Burbidge.etal.1957, Cameron.1957, Cowan.etal.1977,Beers.etal.2005, Sneden.etal.2008}. The elements at the extreme heavy side (e.g., Th and U) are only produced in $r$-process nucleosynthesis \citep{Roederer.etal.2018}. We have a fair understanding of the $s$-process production sites, but the astrophysical sites for $r$-process are still unclear \citep{Saraf.etal.2023,Saraf.et.al.2025}. A complete picture of GCE requires the proper understanding of $r$-process nucleosynthesis.

The primary sites of $s$-process nucleosynthesis are asymptotic giant branch (AGB) stars \citep{Herwig.2005, Campbell.etal.2008, Bisterzo.etal.2010, Doherty.etal.2015}, helium core-burning massive stars \citep{Truran.Iben.1977, Prantzos.etal.1990}, carbon core-burning massive stars \citep{Arnett.etal.1985, Langer.etal.1986, Arcoragi.etal.1991}, and carbon shell-burning stars \citep{Raiteri.etal.1991}. For the $r$-process, we have several proposed sites based on theoretical modeling. These sites include the prompt explosion of low-mass stars \citep{Wheeler.etal.1998,Sumiyoshi.etal.2001,Wanajo.etal.2003}, neutrino driven winds in Type-II supernovae \citep{Woosley.etal.1992,Takahashi.etal.1994, Arcones.etal.2007,Wanajo.etal.2018}, neutron star mergers, neutron star-black hole mergers \citep{Lattimer.etal.1974, Symbalisty.etal.1982, Meyer.etal.1989, Freiburghaus.etal.1999, Goriely.etal.2011, Rosswog.etal.2014,Bovard.etal.2017}, and magnetorotational supernovae and collapsars \citep{Woosley.etal.1993,Nagataki.etal.2007,Fujimoto.etal.2008,Siegel.etal.2018,Siegel.etal.2019}. Recently, the detection of a kilonova event from a neutron star merger supports this as a promising site for $r$-process nucleosynthesis \citep{Arcavi.etal.2017,Smartt.etal.2017, Tanvir.etal.2017,Chornock.etal.2017,Drout.etal.2017,Metzger.etal.2017, Shappee.etal.2017, Tanaka.etal.2017,Villar.etal.2017, Watson.etal.2019}. However, the time delay of neutron star merger may raise issues for the $r$-process enrichment in the early Galaxy.

For a systematic study of $r$-process enrichment, $r$-process stars are divided into three subclasses \citep{Beers.etal.2005, Sneden.etal.2000, Travagilo.etal.2004}:
\begin{align}
\text{$r$-I:} &\quad +0.3 \leq [\text{Eu}/\text{Fe}] \leq +1.0,\quad [\text{Ba}/\text{Eu}] < 0 \notag \\
\text{$r$-II:} &\quad [\text{Eu}/\text{Fe}] > +1.0,\quad [\text{Ba}/\text{Eu}] < 0 \notag \\
\text{limited-$r$:} &\quad [\text{Eu}/\text{Fe}] < +0.3,\quad [\text{Sr}/\text{Ba}] > +0.5, \notag \\
&\quad [\text{Sr}/\text{Eu}] > 0.0 \notag 
\end{align}
This classification is based mainly on the $r$-process element Eu, which is relatively easy to measure in optical spectra. The limit for the Ba abundance is also included to avoid any stars with a dominant contribution from the $s$-process. \cite{Holmbeck.etal.2020} updated this classification using large sample of $r$-process-enhanced (RPE) stars from R-Process Alliance (RPA) and suggested [Eu/Fe] = 0.7 as a separating line between $r$-I and $r$-II subclasses.

Observations of RPE stars provide valuable insights into the astrophysical conditions responsible for the $r$-process. A better understanding requires both chemical and kinematic details. Many studies have recently tried to sort out their chemical peculiarities. However, studies focusing on kinematic aspects are scarce in the literature. 

There have been a number of works attempting to link the kinematics and the chemical peculiarity of stars in order to reveal the chemical enrichment history of the Milky Way. \cite{Navarro.etal.2011} separated thin and thick disk components using abundances of $\alpha$, Fe and $r$-process element Eu. In the [($\alpha$ + Eu)/Fe]$-$[Fe/H] plane, the [($\alpha$ + Eu)/Fe] $=$ 0.2 line clearly separates thin disk and thick disk stars. \cite{Roederer.etal.2018} calculated orbits of 35 highly RPE stars ([Eu/Fe] $\ge$ 0.7) and found that all of these show halo-like kinematics. Using a large sample of nearly 1500 metal-poor stars, \cite{Limberg.etal.2021} identified 38 dynamically tagged groups (DTGs) in energy-action 4D phase space. Several of these groups were previously known streams or substructures. Various RPE stars show similar dynamics to some DTGs indicating a possible origin of RPE stars in dwarf galaxies. Similar DTGs have been identified in \cite{Shank.etal.2022Feb, Shank.etal.2022Aug} for a significantly larger sample of metal-poor stars and discuss their association with known Galactic substructures. \cite{Gudin.etal.2021} identified 30 DTGs in a sample of $r$-process-enhanced metal-poor stars. Although these groups were found based only on kinematics, they show similar chemical signatures, indicating a common chemical evolution history. Subsequently, \cite{Shank.etal.2023} performed a similar analysis for a large sample of RPE stars and they could identify 36 DTGs, including previously known groups. These studies indicate an origin of RPE stars in external systems which are later disrupted and accreted into the Milky Way.

Both spectroscopy and kinematics are important to investigate the site(s) of $r$-process nucleosynthesis 
 \citep{Saraf.etal.2024, Saraf.etal.2024.EPJ}. In this paper, we will combine the spectroscopic and kinematic information of RPE stars to investigate the environmental conditions where the $r$-process occurred and how it enriched the Galaxy with heavy elements. Earlier studies primarily focused on dynamically-tagged groups (DTGs) of stars and the association of RPE stars with these DTGs. Our aim is to separate the RPE stars into different Galactic components (bulge, disk, and halo) and study their chemical peculiarities.

This paper is organized as follows. Section~\ref{sec:introduction} offers an overview of the topic along with a summary of previous research. Section~\ref{sec:sample_selection} outlines the criteria and process used to select data from the literature. The methodology employed is detailed in Section~\ref{sec:orbit_calculation}. Section~\ref{sec:Results} presents the findings of the study, and Section~\ref{sec:conclusions} concludes with a summary of the main insights.

\section{Sample, Orbits, and the Galactic components}

\subsection{Sample selection}
\label{sec:sample_selection}
We have compiled data on RPE stars from \cite{Gudin.etal.2021} and \cite{Shank.etal.2023}, who studied 519 and 1,720 RPE stars, respectively. The former sample includes well-studied RPE stars, while the latter extended their sample by incorporating 1194 RPE stars from the GALAH survey \citep{Buder.etal.2021}. These datasets also provide abundances for C, Fe, Sr, Ba, and Eu, along with a flag indicating the reliability of the abundance estimates. We consider only those objects that have a robust detection of all these elements.
For kinematic analysis, we have obtained parallaxes, proper motions and radial velocities of these objects from Gaia data release 3 \citep{Gaia_edr3_paper, Gaia_dr3_paper}. Since the parallaxes reported by Gaia exhibit a systematic bias, we have corrected for this bias by adding 0.026 mas to the observed parallaxes, as recommended by \cite{Huang.etal.2021}. For the present work, we have adopted the updated definition of $r$-process subclasses from \cite{Holmbeck.etal.2020} and considered only carbon-normal stars with ([C/Fe] $\le$ 0.7). This constrain results in a total of 496 RPE stars with Gaia observations.

\subsection{Orbit calculation}
\label{sec:orbit_calculation}
We have performed back orbit integration of our sample of $r$-process stars in the Milky Way Potential \citep[MWPotential:][]{Bovy.2015.galpy} using the \texttt{GALA} code\footnote{http://gala.adrian.pw/en/latest/} \citep{gala_paper1}. \texttt{GALA} uses \texttt{Astropy} for astronomical units and the coordinate systems. It takes the RA, DEC, parallax, proper motion and radial velocity of an object as input to calculate its trajectory in the chosen potential. It returns various orbital parameters of the object, such as position, velocity, energy, angular momentum, eccentricity, apocenter distance and pericenter distance. For our orbit integration, we adopt following reference frame from \cite{Schonrich.etal.2010} and \cite{Bland-Hawthorn.Gerhard.2016}: distance of the Sun from galactic center (R$_{\odot}$) = 8.2~kpc, velocit of Local Standard of Rest (LSR) = 238~km~s$^{-1}$, vertical location of the Sun (z) = 25~pc, velocity of the Sun with respect to LSR (U$_{\odot}$, V$_{\odot}$, W$_{\odot}$) = (11.1, 12.24, 7.25)~km~s$^{-1}$. In the present work, orbital trajectories are calculated for 8~Gyr back in time with a step size of 0.2~Myr. We have also tested our results with a smaller step size of 0.1~Myr. We could estimate the orbital parameters for 472 RPE stars comprising of 306 $r$-I, 123 $r$-II, and 43 limited-$r$ stars. The remaining 24 objects become unbound during orbital integration.

\subsection{Galactic bulge, disk, and halo}
\label{sec:bulge_disk_halo}
The Milky Way possesses morphologically distinct components known as the bulge, the disk, and the stellar halo (or simply the halo). It is not trivial how to associate stars to these morphological components of the Galaxy. The present location of a star in the Milky Way may be deceiving when it comes to the separation of the three components. A star presently in the disk region may not be part of the disk, but belong instead to the bulge or halo, and being simply presently passing through the disk. In our case, this ambiguity works to our advantage: although our sample is confined to the solar neighborhood, the overlap of dynamical populations in this region allows us to probe stars from multiple Galactic components. This enables a broader exploration of the Galaxy’s structure without requiring observations beyond the local volume. Thus we adopted a physical-motivated definition of Galactic components based on the orbital parameters of the stars: apocenter distance ($r_{a}$) and maximum absolute vertical height ($z_{max}$). We have used the following definitions for different galactic components \citep{Goodwin.etal.1998, Bland-Hawthorn.etal.2016, Zoccali.Valenti.2016}:
\newline
bulge: $r_{a} \le 3$~kpc and $z_{max} \le 3$~kpc
\newline
inner disk: $r_{a} > 3$~kpc, $r_{a} \le 15$~kpc and $z_{max} \le 3$~kpc
\newline
outer disk: $r_{a} > 15$~kpc and $z_{max} \le 3$~kpc
\newline
inner halo: $r_{a} > 3$~kpc, $r_{a} \le 15$~kpc and $z_{max} > 3$~kpc
\newline
outer halo: $r_{a} > 15$~kpc and $z_{max} > 3$~kpc

Figure~\ref{fig:different_components} graphically represents three populations of RPE stars in these five Galactic components separated by the horizontal and vertical lines.

\section{Results}
\label{sec:Results}

\subsection{Orbit based association of RPE stars to Galactic components}
\label{sec:RPE_in_galactic_components}

\begin{figure}
	\centering
	\includegraphics[width=\columnwidth]{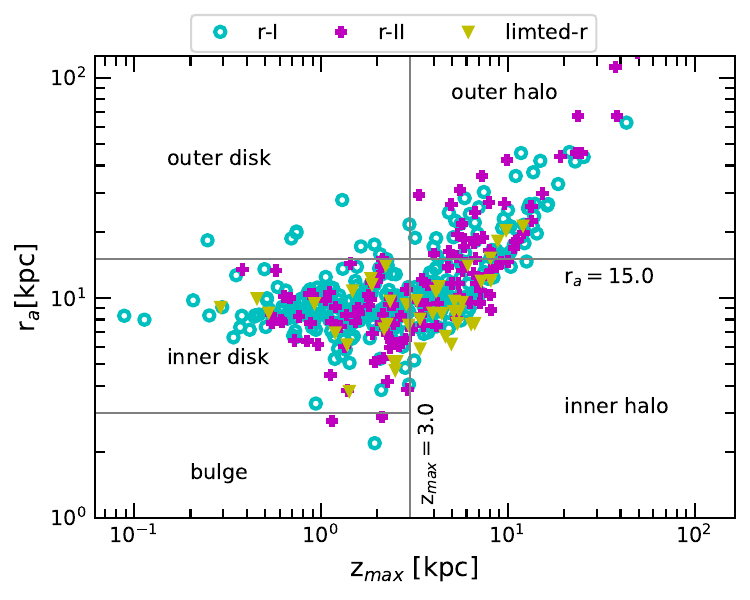}
    \caption{Bulge, disk, and halo classification: The apocenter distance (r$_{a}$) as a function of maximum vertical height (z$_{max}$) of the stellar orbits. For better visibility, we have set the axes scale to log-scale. The vertical black line at z$_{max}=3$ kpc and horizontal black lines at r$_{a}=3$ kpc and r$_{a}=15$ kpc separate RPE stars into five different Galactic components namely, bulge, inner disk, inner halo, outer disk, and outer halo. Open cyan circles show $r$-I stars, $r$-II stars are represented by magenta crosses, and limited-$r$ stars are displayed with yellow down triangles.}
    \label{fig:different_components}
\end{figure}

First, we measured the average apocenter distance (r$_{a}$) and the average maximum absolute vertical height (z$_{max}$) of the stellar orbits during the evolution of 8~Gyr to segregate the stars associated with bulge, disk and halo components. In Figure \ref{fig:different_components}, we have shown $r_{a}$ as a function of $z_{max}$ to separate the RPE stars into different Galactic components using the definitions mentioned in Section~\ref{sec:bulge_disk_halo}. It shows five rectangular regions for bulge, inner disk, inner halo, outer disk, and outer halo stars. The three subclasses of RPE stars $r$-I, $r$-II, and limited-$r$ are respectively represented by open cyan circles, magenta crosses, and yellow down triangles.

Using this orbit-based classification of Galactic components, we found 169 $r$-I, 56 $r$-II, and 19 limited-$r$ are in the inner disk comprising $\sim51\%$ of all the stars. The outer disk has 10 $r$-I and 1 $r$-II star making $\sim2\%$ contribution. There is no limited-$r$ star detected in outer disk. The inner halo shows 76 $r$-I, 28 $r$-II, and 19 limited-$r$ stars consisting of $\sim26\%$. The outer halo has 50 $r$-I, 36 $r$-II, and 5 limited-$r$ stars that contribute to the rest $\sim20\%$. There are only three stars in the bulge region: two $r$-II stars and one $r$-I star. One of the example of disk and halo star are shown in Figure~\ref{fig:disk_and_halo_orbits} in Appendix.

\subsection{Galactic components in the Toomre diagram}
\label{sec:toomre_diagram}

\begin{figure}
	\centering
	\includegraphics[width=\columnwidth]{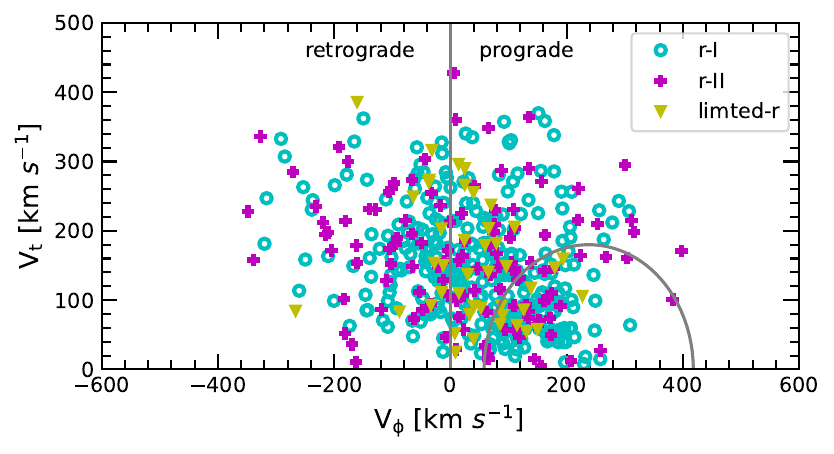}
    \includegraphics[width=\columnwidth]{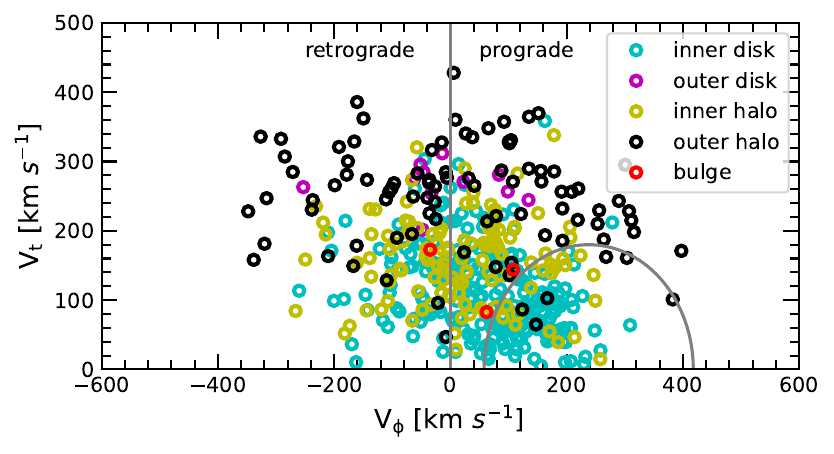}
    \caption{Top panel: Toomre diagram for RPE stars, where the large circle with radius 180 km s$^{-1}$ around LSR is used to separate the disk stars from halo stars and the black vertical line separates the stars on retrograde and prograde motions. Open cyan circles show $r$-I stars, $r$-II stars are represented by magenta crosses, and limited-$r$ stars are displayed with yellow down triangles. Bottom panel: Toomre diagram for $r$-process-enhanced stars color-coded with different galactic components as defined in Section~\ref{sec:bulge_disk_halo}. Colors of different Galactic components are shown in the legend.}
    \label{fig:toomre_diagram}
\end{figure}

The Toomre diagram (the plot of velocity in the plane perpendicular to the Galactic disk as a function of velocity in the Galactic plane) is commonly used to separate the disk and halo stars. It is primarily based on the present day Galactic space velocities (U, V, W), where U is velocity toward the Galactic center, V is azimuthal velocity along the Galactic rotation, and W is toward North Galactic Pole (NGP). We have calculated the present day Galactic space velocities of the sample stars taking the motion of the Sun with respect to LSR from \cite{Schonrich.etal.2010}, i.e., (U$_{\odot}$, V$_{\odot}$, W$_{\odot}$) = (11.1, 12.24, 7.25) km s$^{-1}$.

In Figure~\ref{fig:toomre_diagram}, we show the Toomre diagram for our sample. Here, the x-axis shows azimuthal velocity (V$_{\rm \phi}$ = V) and the y-axis shows transverse velocity (V$_{\rm t}$ = $\sqrt{U^{2} + W^{2}}$) of stars. The top panel shows the distribution of different $r$-process subclasses and the bottom panel shows their association with different Galactic components defined in Section~\ref{sec:bulge_disk_halo}. The gray-color semi-circle is commonly used as the boundary between disk and halo stars. Stars within the semi-circle are associated with the disk and those outside the semicircle are associated with the halo. According to this definition the majority of $r$-process-enhanced stars are part of the halo. The vertical line at V $=$ 0 separates prograde (V $>$ 0: moving in the direction of Galactic rotation) and retrograde (V $<$ 0: moving in the opposite direction of Galactic rotation) stars. We found $\sim 65\%$ prograde (201 $r$-I, 73 $r$-II, and 31 limited-$r$) and $\sim 35\%$ retrograde (105 $r$-I, 50 $r$-II, and 12 limited-$r$) orbits in all $r$-process stars.

\begin{figure}
	\centering
	\includegraphics[width=\columnwidth]{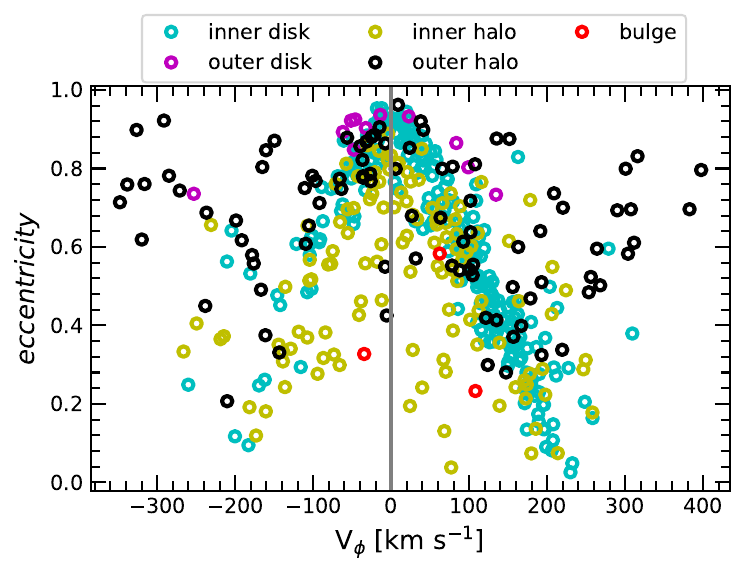}
    \caption{Orbital eccentricity of $r$-process-enhanced stars as a function of azimuthal velocity. Stars are color-coded with different galactic components as defined in Section~\ref{sec:bulge_disk_halo}. Colors of different Galactic components are shown in the legend at the top of the figure.}
    \label{fig:ecc_V_Gal_Comp}
\end{figure}

In the bottom panel of Figure~\ref{fig:toomre_diagram}, our orbital parameter-based definition of Galactic components shows that the Toomre diagram does not separates disk and halo components properly. Although it shows most halo stars are outside of the semi-circle, there are also plenty of disk stars outside the semi-circle. One may argue that these retrograde disk stars (V$_{\phi}<$ 0) should be halo stars. However, our orbital calculations show that they have remained confined to the disk over the past 8~Gyr. Therefore, they are more likely to belong to the counter-rotating disk component of the Galaxy. To investigate the main cause of this discrepancy, we plot orbital eccentricity as a function of azimuthal velocity in Figure~\ref{fig:ecc_V_Gal_Comp}. Orbital eccentricity is defined as $e= (r_{a} - r_{p})/(r_{a} + r_{p})$, where $r_{a}$ is apocenter distance and $r_{p}$ is pericenter distance of the orbit. This parameter is useful to quantify the shape of the orbit. Its value ranges from 0 for a circular orbit to 1 for an eccentric orbit. Figure~\ref{fig:ecc_V_Gal_Comp} shows that a large fraction of disk stars are on retrograde orbits (V$_{\phi}<$ 0). Also, if we see the eccentricity of prograde orbits, it shows a clear decreasing trend with increasing azimuthal velocity. Highly eccentric orbits ($e$ very close to 1) have higher radial velocity components that makes transverse velocity higher. Both retrograde motion and eccentric orbits move star outside the disk-halo separating semi-circle. It confirms that the Toomre diagram does not provide a good diagnostic to separate disk and halo stars when orbits are highly eccentric and/or retrograde. Thus, we recommend using orbital parameter-based classification for the stars to associate them with the Galactic components.

\subsection{In-situ and ex-situ origin of RPE stars}
\label{sec:in-situ_ex-situ}

\begin{figure}
	\centering
	\includegraphics[width=\columnwidth]{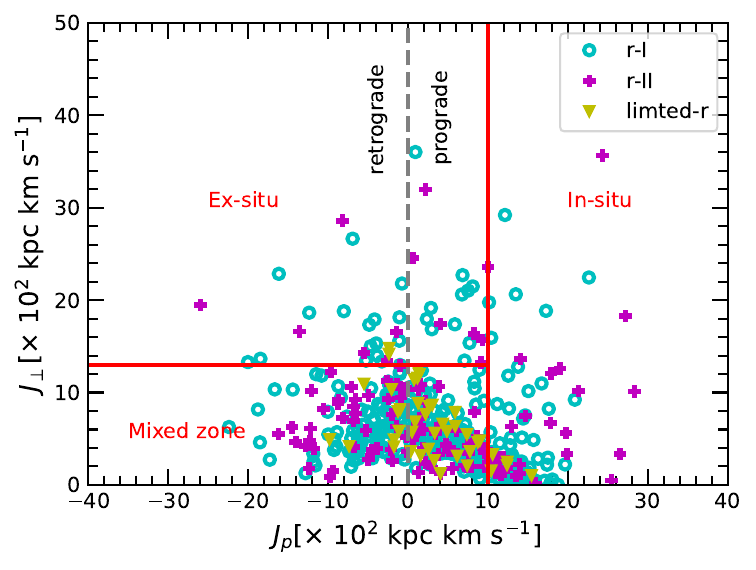}
    \caption{This is the specific angular momentum diagram for RPE stars, where the vertical black dotted line separates the stars with retrograde and prograde motions. The vertical red line at $J_{p}=1000$ kpc km s$^{-1}$, and horizontal red line at $J_{\perp}=1300$ kpc km s$^{-1}$ separate in-situ, ex-situ and mixed stars. Open cyan circles show $r$-I stars, $r$-II stars are represented by magenta crosses, and limited-$r$ stars are displayed with yellow down triangles.}
    \label{fig:specific_angmom}
\end{figure}

RPE stars have also been observed in nearby dwarf satellite galaxies such as Carina, Draco, Fornax, Ursa Minor (UMi), and Sculptor. The discovery of highly RPE stars in ultrafaint dwarf galaxy Reticulum II (Ret II) received large attention. There were seven $r$-II stars in nine studied stars \cite{Ji.etal.2016}. The evidence of RPE stars in satellite galaxies and in Milky Way halo suggest that RPE stars may have formed in external systems which are disrupted and accreted into Milky Way \cite{Ji.etal.2016, Roederer.etal.2018}. \cite{Brauer.etal.2019} suggested that half of the $r$-II stars in the Milky Way may have originated in now-destroyed dwarf galaxies.

To investigate the in-situ (formed in the Galaxy) and the ex-situ (accreted from other stellar systems) formation history of RPE stars, we plot the total radial and vertical action ($J_{\perp} = \sqrt{J_{r}^{2} + J_{z}^{2}}$) as a function of azimuth action ($J_{p}$) in Figure~\ref{fig:specific_angmom}. We prefer using actions over angular momenta because angular momenta about the axis in the Galactic plane vary significantly during the orbit of a star in the Milky Way Potential. Following \cite{Di_Matteo.etal.2020}, we categorized stars into in-situ, ex-situ, and mixed-zone origin. Stars with $J_{p}>1000$ kpc km s$^{-1}$ are formed in-situ. Stars with $J_{p}<1000$ kpc km  s$^{-1}$ and $J_{\perp}>1300$ kpc km  s$^{-1}$ are formed ex-situ. Stars with $J_{p}<1000$ kpc km  s$^{-1}$ and $J_{\perp}<1300$ kpc km s$^{-1}$ are formed either in-situ or ex-situ and thus kept in the mixed-zone. Our analysis showed that $\sim 24\%$ (80 $r$-I, 28 $r$-II, and 5 limited-$r$) are formed in-situ, $\sim 10\%$ (28 $r$-I, 17 $r$-II, and 3 limited-$r$) are formed ex-situ, and $\sim 66\%$ (198 $r$-I, 78 $r$-II, and 35 limited-$r$) are in the mixed-zone.

It is evident that about 3/4 of RPE stars are in the mixed-zone. An individual $r$-process subclass also roughly shows a 3/4 fraction in the mixed-zone. Based on our classification of Galactic components (for brevity, the figure is not shown here), we found that all ex-situ stars belong to the halo region. Since, all the outer disk stars are on highly eccentric orbits as can be seen in Figure~\ref{fig:ecc_V_Gal_Comp}, they lie in the mixed zone.

\subsection{R-process enrichment of Galactic components}
\label{RPE_in_galactic_components}

\begin{figure}
    \centering
    \includegraphics[width=\columnwidth]{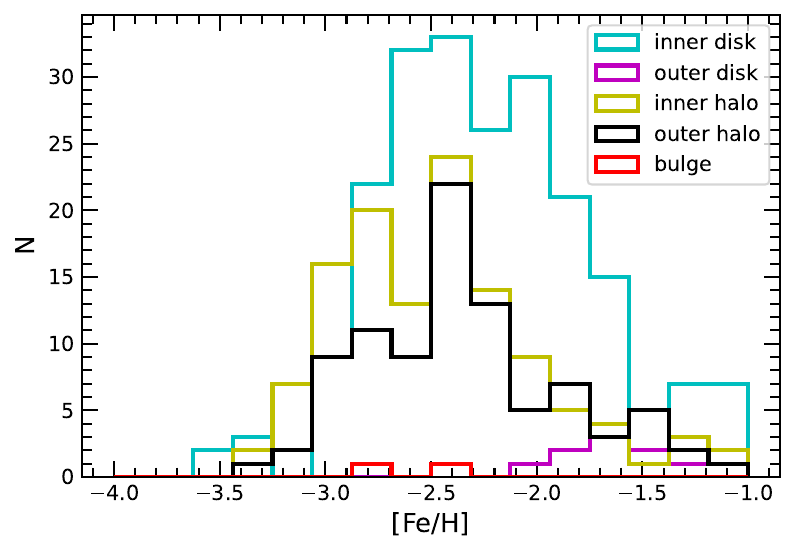}
    \caption{Metallicity ([Fe/H]) distribution of RPE stars in different Galactic components.}
    \label{fig:FeH_Gal_Comp}
\end{figure}

As a first step to understand the chemical evolution of RPE stars in different Galactic components, we calculate the number distribution of RPE stars as a function of metallicity. Figure~\ref{fig:FeH_Gal_Comp} shows the metallicity histogram of sample RPE stars in bulge, inner disk, inner halo, outer disk, and outer halo. Inner disk shows bi-modality in the metallicity distribution peaking around [Fe/H] = $-2.5$ and $-2.0$ dex. Outer disk RPE stars are lying toward the metal-rich end of the distribution, whereas inner and outer halos lie toward metal-poor end. Additionally, the peaks of the inner and outer halos coincide with the first peak ([Fe/H]=$-2.5$) of the inner disk distribution. There is also a hint of bi-modality in inner and outer halos. The bi-modal metallicity distribution of inner disk and halo could be the result of different formation mechanisms or assembly time.

\begin{figure}
    \centering
    \includegraphics[width=\columnwidth]{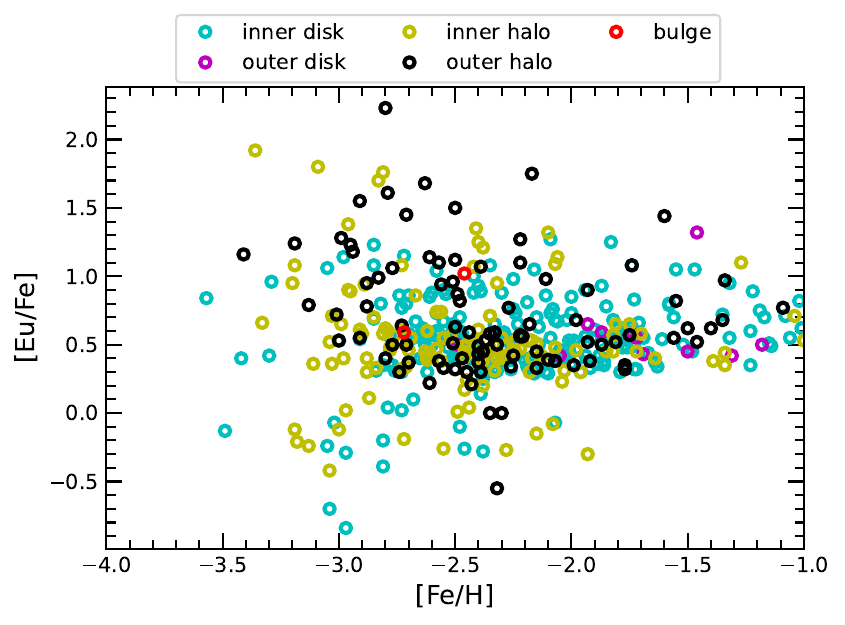}
    \caption{Eu abundance of RPE stars as a function of metallicity ([Fe/H]) for different Galactic components.}
    \label{fig:EuFe_FeH_Gal_Comp}
\end{figure}

Figure \ref{fig:EuFe_FeH_Gal_Comp} shows the [Eu/Fe] as a function of [Fe/H] for RPE stars in different Galactic components. It is clear that inner disk, inner halo and outer halo exhibit similar abundance trends with metallicity. All these components show large scatter at low metallicity, except the outer disk stars. The stars in outer disk region show nearly constant [Eu/Fe] abundance. Limited-$r$ stars are generally close to the inner disk and halo. Similar distribution is observed for the Ba abundance as well, but not shown for brevity.

\begin{figure}
    \centering
    \includegraphics[width=\columnwidth]{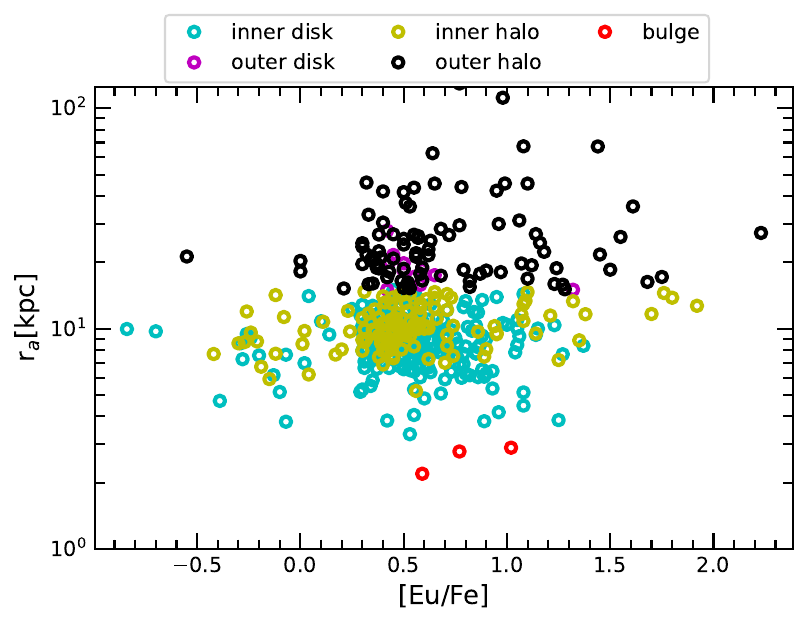}
    \caption{Apocenter distance of RPE stars as a function of metallicity for different Galactic components.}
    \label{fig:ra_vs_EuFe_Gal_Comp}
\end{figure}

We have also calculated the [Eu/Fe] distribution in Galactic components as a function of their apocenter distances and shown in Figure~\ref{fig:ra_vs_EuFe_Gal_Comp}. RPE stars in the inner disk, inner halo and outer halo show a range of Eu abundances. However, outer disk stars are clustered in a smaller region. In Figure~\ref{fig:ecc_V_Gal_Comp}, we found outer disk stars are on highly eccentric orbits and the majority of them are on retrograde motion. This indicates that RPE stars in the outer disk may have born in a single event/environment. The above conclusions are tentative given the small number of stars in each of the categories.







\section{Conclusions}
\label{sec:conclusions}
We have compiled a sample of 472 metal-poor $r$-process-enhanced (RPE) stars from the literature for their kinematic study. It is comprised of 43 limited-$r$, 306 $r$-I, and 123 $r$-II stars. We perform back orbit integration for 8~Gyr to estimate their orbital parameters. We associate these objects to bulge, inner disk, inner halo, outer disk, and outer halo components of the Galaxy using apocenter distance and maximum vertical height of the orbit. Following are our findings.

\begin{itemize}
    \item
    Our orbital parameter based classification of RPE stars result in 3 bulge, 244 inner disk, 123 inner halo, 11 outer disk, and  91 outer halo stars. Thus, our classification shows nearly similar fraction of RPE stars in the disk (inner + outer $\approx$~52\%) and in the halo (inner + outer $\approx$~48\%).
    \item
    We showed that the Toomre diagram based classification of disk and halo stars is not very useful for highly eccentric and/or retrograde orbits. We recommend to use physically motivated orbital-based definitions for separating bulge, disk, and halo stars.
    \item
    Using specific angular momenta of the orbits, we found that all the most probably accreted (ex-situ) RPE stars belong to the halo region. However, around 3/4 RPE stars are in a mixed zone. This majority of stars in the mixed zone complicates the understanding of in-situ and ex-situ origin.
    \item
    The RPE stars in the outer disk region show clustering in [Eu/Fe]-r$_{a}$ plane. These stars are on highly eccentric orbits, with the majority in retrograde motion. It indicates they may have come from the same $r$-process event.
    \item 
    The inner disk, inner halo and outer halo stars do not show a significant difference in the abundance trends of n-capture elements.  Limited-$r$ stars are found more closure to the inner disk and halo compared to $r$-I and $r$-II stars.
\end{itemize}

Our analysis is based on a relatively small sample of metal-poor RPE stars. A comprehensive analysis of $r$-process enrichment requires a large sample of RPE stars with detailed kinematics. 


\appendix
\section{Example orbits of disk and halo stars}

\begin{figure*}[!ht]
     \centering
     \includegraphics[width=0.9\textwidth]{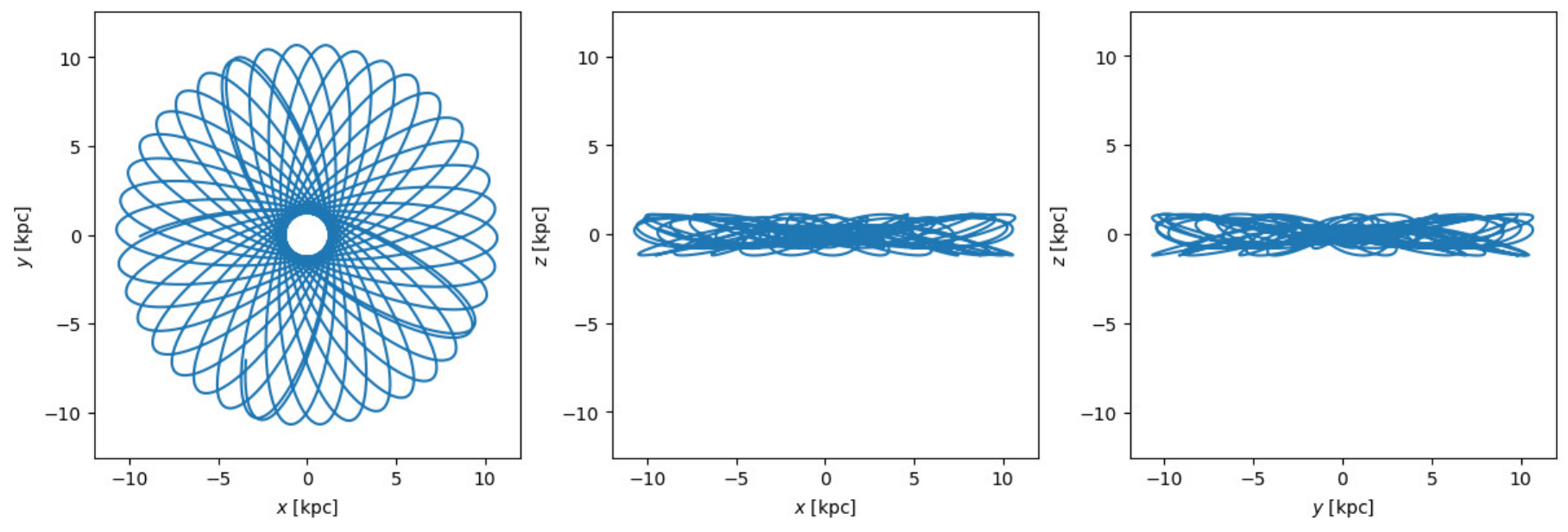}
     \includegraphics[width=0.9\textwidth]{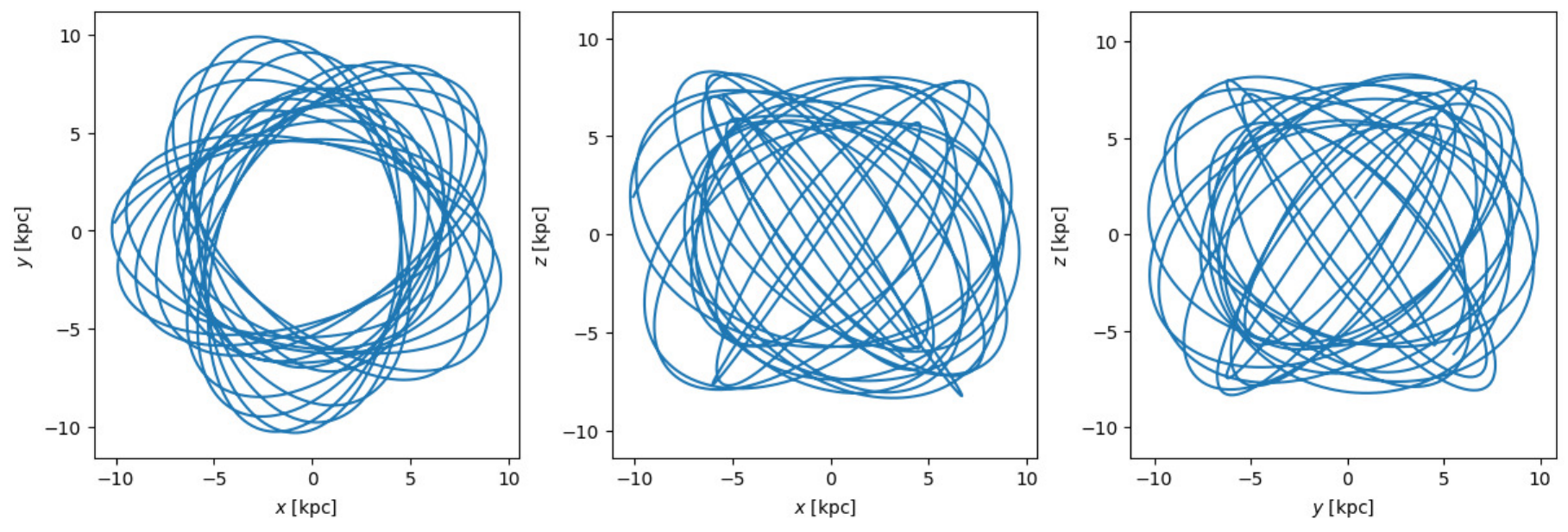}
     \caption{Example orbits of two stars in in the Milky Way potential (MW Potential). Top panel shows the star orbiting in disk-plane and bottom panel shows the star orbiting in the stellar halo.}
     \label{fig:disk_and_halo_orbits}
 \end{figure*}
 
For illustration, in Figure~\ref{fig:disk_and_halo_orbits}, we have the orbits of stars in disk and halo of the Milky Way. The first row panels show a star orbiting in the Galactic plane and the second row panel shows a star in the halo. In each row, left column displays the projected orbit in the disk plane. The middle and right columns represent the side-on view of the orbiting stars. The disk stars always remain very close to the Galactic plane, whereas halo stars reach farther than 3~kpc from the Galactic plane as evident from middle and right columns of the figure.
%

\section*{Acknowledgements}
This work has made use of data from the European Space Agency (ESA) mission {\it Gaia} (\url{https://www.cosmos.esa.int/gaia}), processed by the {\it Gaia} Data Processing and Analysis Consortium (DPAC, \url{https://www.cosmos.esa.int/web/gaia/dpac/consortium}). Funding for the DPAC has been provided by national institutions, in particular the institutions participating in the {\it Gaia} Multilateral Agreement. This work made use of Astropy:\footnote{http://www.astropy.org} a community-developed core Python package and an ecosystem of tools and resources for astronomy \citep{astropy:2013, astropy:2018, astropy:2022}. This publication makes use of VOSA, developed under the Spanish Virtual Observatory (https://svo.cab.inta-csic.es) project funded by MCIN/AEI/10.13039/501100011033/ through grant PID2020-112949GB-I00. This work make use of Matplotlib \citep{Matplotlib.2007}, NumPy \citep{Numpy.2020}, and Pandas \citep{Pandas.paper.2010, Pandas.software.2022}. The work at PRL is supported by Department of Space.
CAP is thankful for funding from the Spanish government through grants AYA2014-56359-P, AYA2017-86389-P and PID2020-117493GB-100. 
\vspace{-1em}



\end{document}